\documentclass[aps,prb,twocolumn,showpacs]{revtex4}

\usepackage{epsfig}

\usepackage{afterpage}

\usepackage{epsfig,amsmath}

\usepackage{color}

\usepackage{amsmath}

\usepackage{amssymb}

\usepackage[latin1]{input enc}

\usepackage{graphics}

\usepackage{graphicx}

\usepackage{subfigure}

\usepackage{psfrag}

\begin{document}

\newcommand{\nn}{\nonumber}

\newcommand{\bb}{\begin{eqnarray}}

\newcommand{\ee}{\end{eqnarray}}

\newcommand{\Vol}{{\cal A}}

\newcommand{\sign}{{\rm sign}}

\newcommand{\LL}{\textsf{L} }

\newcommand{\Ar}{{\cal A}_0}

\newcommand{\ar}{\sigma}

\newcommand{\ff}{\frac{1}{2}}

\newcommand{\Tc}{\tilde T}

\newcommand{\Bc}{\tilde B}

\newcommand{\Ec}{\tilde E}

\newcommand{\amp}{A_1}

\newcommand{\EF}{\mu_0}

\newcommand{\bin}[2]{C_{#1}^{#2}}
\newcommand{\coef}{{\cal S}}

\title{\textbf{Random walks and magnetic oscillations in
compensated metals}}

\author{Jean-Yves Fortin$^1$}\email{fortin@lpm.u-nancy.fr}

\author{Alain Audouard$^2$}\email{audouard@lncmp.org}

\affiliation{ $^1$Institut Jean Lamour
D\'epartement de Physique de la Mati\`ere et des Mat\'eriaux,
Groupe de Physique Statistique,
CNRS - Nancy-Universit\'e BP 70239
F-54506 Vandoeuvre les Nancy Cedex, France\\
$^2$Laboratoire National des Champs Magn\'etiques Intenses (UPR
3228 CNRS, INSA, UJF, UPS) 143 avenue de Rangueil, F-31400
Toulouse, France}

\date{\today}

\begin{abstract}

The field- and temperature-dependent de Haas-van Alphen
oscillations spectrum is studied for an ideal two-dimensional
compensated metal whose Fermi surface is made of a linear chain of
successive  orbits with electron and hole character, coupled by
magnetic breakdown. We show that the first harmonics amplitude can
be accurately evaluated on the basis of the Lifshits-Kosevich (LK) formula by
considering a set of random walks on the orbit network, in
agreement with the numerical resolution of semi-classical
equations. Oppositely, the second harmonics amplitude does not
follow the LK behavior and vanishes at a critical value of the
field-to-temperature ratio which depends explicitly on the relative 
value between the hole and electron effective masses.

\end{abstract}

\pacs{71.18.+y,71.20.Rv,74.70.Kn}

\maketitle

\section{Introduction}

Due to their rather simple Fermi surface (FS), organic metals
provide a powerful playground for the investigation of quantum
oscillation physics. In that respect, the most well known example
is provided by $\kappa$-(BEDT-TTF)$_2$Cu(NCS)$_2$ which can be
regarded as the experimental realization of the FS considered by
Pippard in the early sixties for his model\cite{Pi62}. Such a FS,
composed of closed hole orbits and quasi-one dimensional sheets
coupled by magnetic breakdown (MB), yields quantum oscillations
spectra with numerous frequency combinations that cannot be
accounted for by the semiclassical model of Falicov and
Stachowiak\cite{Fa66}. This phenomenon which is attributed, for
this kind of FS, to either the formation of Landau bands or (and)
the oscillation of the chemical potential in magnetic field has
raised a great interest\cite{Al96,Sa97,Fo98,Gv02,Ch02,Ki02,Fo05}.
Nevertheless, the FS of numerous organic metals is composed of
compensated electron- and hole-type closed orbits\cite{Ro04},
yielding many frequency combinations as well, as far as
Shubnikov-de Haas oscillations are concerned\cite{Pr02,Vi03,Au05}.
In the case of a FS composed of two compensated orbits coupled to
each other through MB but isolated from the other orbits outside
the first Brillouin zone (FBZ), it has been shown that the
oscillations of the chemical potential can be strongly damped
\cite{Fo08} which could account for the absence of frequency
combinations reported in the de Haas-van Alphen (dHvA) spectra of
two-dimensional (2D) networks of compensated orbits in fields up
to 28 T \cite{Au05}. However, the FS considered in Ref.
\onlinecite{Fo08} which, to our knowledge, has no counterpart
among the compounds synthesized up to now, does not provide a
network of orbits and, therefore, do not yield Landau bands in
magnetic field.

\begin{figure}                                                      
\centering \resizebox{0.8\columnwidth}{!}
{\includegraphics*{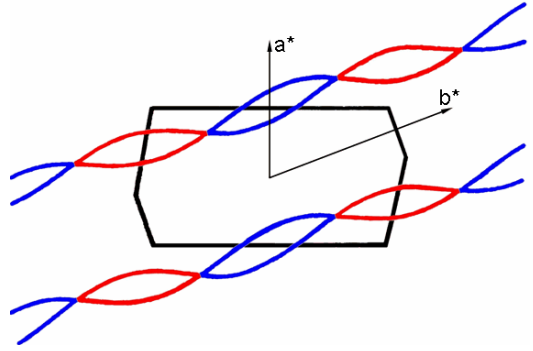}}
\caption{\label{fig_SF_NO3}(color online) Calculated Fermi surface
of the Bechgaard salt (TMTSF)$_2$NO$_3$ in the temperature range
below the anion ordering and above the spin density wave
condensation, according to Ref. \onlinecite{Ka05}. Solid blue and
red lines are compensated hole and electron orbits, respectively.}
\end{figure}

The aim of this paper is to explore the field and temperature
dependence of the dHvA oscillations spectra of an ideal 2D metal
whose FS achieves a linear chain of successive electron- and
hole-type compensated orbits. Such a topology can be relevant in
compounds for which the FS originates from an orbit with an area
equal to that of the FBZ and coming close to the FBZ boundary
along one direction, as it has been predicted for
(BEDT-TTF)$_4$NH$_4$[Fe(C$_2$O$_4$)$_3$]\cite{Pr03}. In this case,
a large MB gap is observed at this point and a linear chain of
successive electron-hole tubes separated with a smaller MB gap can
be observed. Analogous topology is also realized in the Bechgaard
salt (TMTSF)$_2$NO$_3$ in the temperature range in-between the
anion ordering temperature and the spin density wave condensation
\cite{Ka05}. We will focus on the field- and temperature-dependent
amplitude of the first and second harmonics $A_1$ and $A_2$ of the
oscillatory part of the magnetization $M_{osc}$ derived for such
FS topology, which can be expanded as $M_{osc}$ = $\sum_i A_i
\sin(2\pi iF_0/b)$, where the fundamental frequency of the problem
$F_0$  and the dimensionless magnetic field $b$ are specified
in the next section.

\section[model]{Model}

\begin{figure}
\centering
\resizebox{1\columnwidth}{!}{\includegraphics*{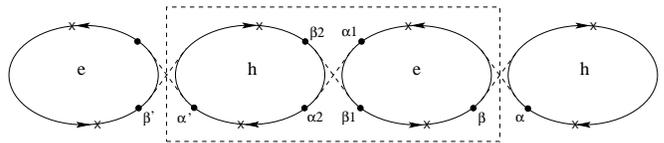}}
\caption{\label{model} One dimensional chain of coupled electron
and hole orbits. The Brillouin zone is delimited by the dotted
lines. Amplitudes are chosen at specific points on the trajectory
(see text).}
\end{figure}

We consider a 2D metal whose electronic structure consists of two
parabolic bands with hole- and electron-character yielding  a
periodic array of compensated orbits (see Fig. \ref{model}). The
bottom of the electron band is set at zero energy while the top of
the hole band is at $\Delta>0$, with the possibility for the
quasiparticle to tunnel through a gap between two successive
orbits by MB. The total number of quasiparticles in the system is
constant and, to lower the total energy at zero field and zero
temperature, the area of the hole-band at the Fermi surface should be
equal to the area of the electron-band which accounts for
compensation. Indeed, if the hole band were completely filled, the
zero-field energy would be higher than if one part of the
quasiparticles were transferred from the hole-band to the
electron-band. In the following, the effective masses linked to
the two bands, $m^*_e$ and $m^*_h$, can be different. As in Refs.
\onlinecite{Fo05,Fo08}, the dimensionless field $b$ and
temperature $t$ are given by $b= B/\tilde{B}$ and $t =
T/\tilde{T}$, respectively, where $\tilde{B} = h/eA_0$,
$\tilde{T}$= $\tilde{E}/k_B$, and $A_0$ is the unit cell area. The
effective masses and energies are expressed in free electron mass
$m_0$ units and in units of $\tilde{E}$ = 2$\pi\hbar^2m_0A_0$,
respectively. Unit cell area of most organic metals is in the
range 100-200 ${\rm \AA^2}$ yielding $\tilde{B}$ and $\tilde{T}$
values of few thousands of Tesla and Kelvin, respectively.
Therefore, realistic experimental conditions yield small values of
$b$ and $t$ compared to $\tilde{B}$ and $\tilde{T}$, respectively.
On the contrary, the ratio $b/t$, which is the relevant external
parameter for perfect crystals, is given by $b/t =(e\hbar/m_0k_B)$
$B/T$. Its value is close to the $B/T$ ratio achieved in
experiments since $e\hbar/m_0k_B$ $\simeq$ 1.34 KT$^{-1}$. Given
an energy $E$, the areas of the electron- and hole-type surfaces
are respectively $S_e=2\pi m_e^*E$ and $S_h=2\pi m_h^*(\Delta-E)$,
which are both quantized for closed orbits in the Brillouin zone.
The zero field Fermi energy $\EF$ is given by the condition of
compensation $S_e(\EF)=S_h(\EF)$ hence
$\EF=[m_h^{*}/(m_e^{*}+m_h^{*})]\Delta$. The fundamental frequency
of this system is therefore equal to
$F_0=S_e(\EF)/2\pi=S_h(\EF)/2\pi=m_e^*m_h^*\Delta/(m_e^{*}+m_h^{*})$.
 Each Landau level (LL) has a degeneracy $b$ per sample
area. The spectrum of such chain of coupled electron-hole orbits
and the quantization of the energy are determined by
semi-classical and conservation rules of the wave-function
amplitudes at the junctions where the quasiparticles tunnel and
across the boundaries of the Brillouin zone. In particular, the
amplitude of the wave-function at different points of the Fermi
surface (see Fig. \ref{model} for notations) satisfies the following rules: its
phase is proportional
to the area swept by the quasi-particle around the trajectory
divided by $b$. We will note $\sigma_e=S_e/2b$ and
$\sigma_h=S_h/2b$ the phases around half the electron and hole
orbits, respectively. Also, at every junction there is a
probability $ip$ for the quasi-particle to tunnel and a
probability $q$ to be reflected. Finally we add a Maslov index
$i=\sqrt{-1}$ at the vertical extrema of the orbits (cross symbols
on Fig. \ref{model}). With these rules, we can write the relations
between the wave-function amplitudes $\alpha$'s and $\beta$'s:

\bb\nn
\alpha_1&=&i\exp(i\sigma_e)(ip\alpha+q\beta), \\ \nn
\alpha_2&=&ip\alpha_1+q\beta_2,\;\;\alpha'=i\exp(-i\sigma_h)\alpha_2
\\
\beta_1&=&q\alpha_1+ip\beta_2, \\ \nn
\beta_2&=&i\exp(-i\sigma_h)(q\alpha'+ip\beta').
\ee

As boundary conditions, we impose that the
amplitudes across one Brillouin zone are identical up to an arbitrary phase
$\theta$: $\alpha'=\alpha\exp(i\theta)$ and $\beta'=\beta\exp(i\theta)$.
Solving this system of closed equations, we obtain the quantization of the
energy which satisfies the spectrum equation

\bb\nn
& &\Big [
1+q^2\exp(2i\sigma_e)-p^2\exp i(\sigma_e-\sigma_h+\theta)
\Big ]\times
\\ \nn
& &\Big [
1+q^2\exp(-2i\sigma_h)-p^2\exp i(\sigma_e-\sigma_h-\theta)
\Big ]
\\ \nn
&=&
-p^2q^2
\Big [
\exp i(\sigma_e-\sigma_h)+\exp(-2i\sigma_h+i\theta)
\Big ] \times
\\ \label{spectrum}
& &\Big [
\exp i(\sigma_e-\sigma_h)+\exp(2i\sigma_e-i\theta)
\Big ]
\ee

We can distinguish two limiting behaviors. For $q = 1$ (or,
equivalently, $p = 0$), the spectrum reduces to
$[1+\exp(2i\sigma_e)]\times[1+\exp(-2i\sigma_h)]$ = 0 which
corresponds to the discrete LL of two independent electron and
hole orbits: $S_e(E_n)=2\pi b(n+1/2)$ and $S_h(E_n)=2\pi
b(n+1/2)$, with $n$ positive integer. In the opposite case where
$q$ = 0, the spectrum corresponds to non-quantized phases
$\sigma_e(E)-\sigma_h(E)=\pm\theta$ since no closed orbit exists
in such case (open chain) and therefore magnetic oscillations
vanish, leaving the place to a continuum of states. To study
numerically the spectrum given by Eq. (\ref{spectrum}), we first
determine the periodicity in energy of the discrete LLs. For that
purpose, we assume that $m^*_h/m^*_e$=$q_0/p_0$ where $q_0$
and $p_0$ are coprime integers that can be as large as needed to
approximate the ratio of the two masses. It is indeed useful to express this
quantity with a rational number since it allows to consider as necessary only a
finite set of solutions of Eq. (\ref{spectrum}). The minimal
periodicity $T_E$ of the spectrum is, in this case and for a given
phase $\theta$, equal to $T_E=2bp_0/m_e^*$, which is twice the
periodicity calculated for an isolated system made of one hole and
one electron band \cite{Fo08}. The number of solutions inside each
interval of width $T_E$ is found to be equal to $2(p_0+q_0)$ (this
number is conserved when $p$ varies and can be counted exactly for
$p=0$). Given a set of solutions $E_{eh}(q,n,\theta)$,
$n=0,\cdots,2(p_0+q_0)-1$, we introduce a cutoff function
$\varphi_c(E)$ such as $\varphi_c(E)=1$ for $E$ larger than a
characteristic energy $E_c$ and equal to
$\exp[-c(E-E_c)^{2\delta}]$ for $E\le E_c$, where $\delta$ is any
positive integer greater than 1 (we take $\delta=4$ in the
simulations which gives a very smooth cutoff function), and $c$ a positive
parameter determined self-consistently. This
function has the property of preserving the physical features near
the Fermi surface and the GS energy is in particular finite since
now the corresponding hole spectrum is bounded for large and negative energies.
The LL density of states $\rho_c(E)$ takes the following form

\bb\nn
\rho_c(E)&=&\frac{b}{2\pi}\int_0^{2\pi}d\theta\sum_{n=0}^{2(p_0+q_0)-1}
\sum_{k=-\infty } ^ { \infty }
\\
& &\varphi_c(E)
\delta(E-E_{eh}(q,n,\theta)-kT_E)
\ee

Given $E_c$, the positive parameter $c$ is found to be solution of the equation
of conservation at zero temperature:

\bb\nn
N_{eh}&=&\int_{-\infty}^{\EF(b)} dE \rho_c(E)
\\ \nn
&=&
\frac{b}{2\pi}\int_{-\infty}^{\EF(b)} dE
\int_0^{2\pi}d\theta\sum_{n=0}^{p_0+q_0-1}\sum_{k=-\infty}^{\infty}
\\ \label{cons}
& &\varphi_c(E)\delta(E-E_{eh}(q,n,\theta)-kT_E)
\ee

where $N_{eh}$ is, as mentioned before, the total number of quasiparticles in
the Canonical Ensemble. In the numerical simulations, $\EF(b)$ is taken as the
first LL located below $\EF$. We
choose $N_{eh}$, which is arbitrary, as a multiple of
the characteristic zero field energy density $(m_e^*+m_h^*)\EF$ (which is
proportional to the areas of the Fermi surface). We will also choose in the
following $E_c=-2$ and $N_{eh}=8(m_e^*+m_h^*)\EF$, and
10 values of $\theta$ for the integral evaluations.
For each value of the field $b$, the parameter $c$ defined from Eq.
(\ref{cons}) is unique and determined self-consistently. Then
we can compute for example the GS energy $\Delta E_0$

\bb\nn
\Delta E_0&=&\frac{b}{2\pi}\int_{-\infty}^{\EF(b)} dE
\int_0^{2\pi}d\theta
\sum_{n=0}^{p_0+q_0-1}\sum_{k=-\infty}^{\infty}
\\ \label{GSE}
& &E\varphi_c(E)\delta(E-E_{eh}(q,n,\theta)-kT_E)
\ee

and the free energy $\Delta F$

\bb\nn
\Delta F&=&-\frac{tb}{2\pi}\int_{-\infty}^{\infty} dE\int_0^{2\pi}d\theta
\sum_{n=0}^{p_0+q_0-1}\sum_{k=-\infty}^{\infty}\varphi_c(E)
\\ \label{freeE} \nn
& &\log[1+\exp\beta(\mu-E)]\delta(E-E_{eh}(q,n,\theta)-kT_E)
\\
&+&N_{eh}\mu.
\ee

The chemical potential $\mu=\mu(t,b)$ is calculated from Eq. (\ref{freeE}) by
extremizing the free energy $\partial \Delta F/\partial \mu=0$. The
magnetization $M_{osc}=-\partial \Delta F/\partial b=x^2\partial \Delta
F/\partial x$, where $x=1/b$, has to be independent of
the parameter $E_c$ for $E_c$ far away from the chemical potential or at
energies large compared to the Landau gap for a given magnetic field. We have
checked, for different values of $E_c$ with $\Delta=1$, for example
$E_c=-1,-2,-4$, and for a large range of fields, that the resulting
magnetization is stable in this procedure.

\begin{figure}                                                      
\centering \resizebox{0.8\columnwidth}{!}
{\includegraphics*{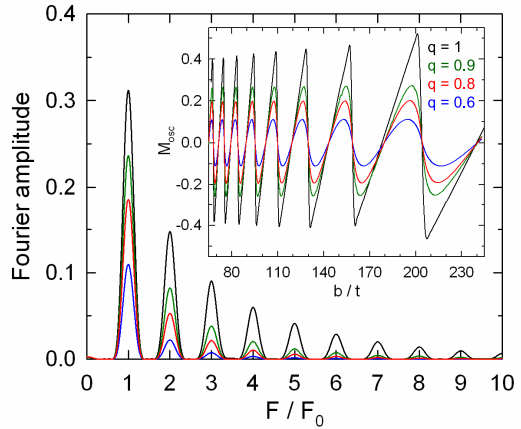}}
\caption{\label{fig_dHvA_TF}(color online) Fourier spectra of the
oscillatory magnetization (displayed in the inset) for various $q$
values. The effective masses are $m_e$ = 1 and $m_h$ = 2.5. $F_0$
is the fundamental frequency (see text).}
\end{figure}

\begin{figure}                                                      
\centering \resizebox{0.8\columnwidth}{!}
{\includegraphics*{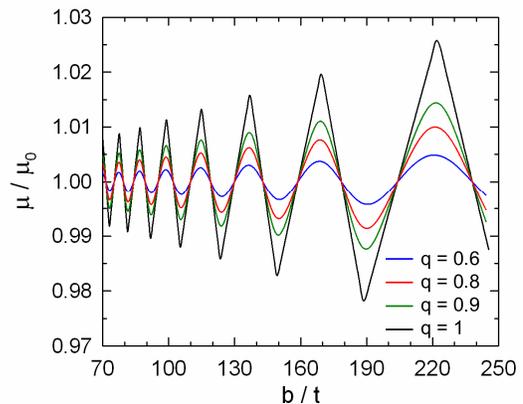}} \caption{\label{fig_mu}(color online)
$b/t$ dependence of the chemical potential normalized to its value
$\EF$ in zero-field for various $q$ values. The effective masses are
$m_e$ = 1 and $m_h$ = 2.5.}
\end{figure}

Examples of field-dependent dHvA and chemical potential
oscillations deduced from the numerical resolution of Eq.
\ref{freeE} are given in Figs. \ref{fig_dHvA_TF} and \ref{fig_mu},
respectively. These data are derived assuming a constant
tunnelling probability $p=\sqrt{1-q^2}$ even though $p$ is
field-dependent. Nevertheless, a qualitative conclusion can be
derived from these data which otherwise will be used in the
following for the study of the Fourier amplitudes: the
oscillations of the chemical potential are weak, owing to the
large $b/t$ range considered, and decrease as $q$ decreases. This
point which is in agreement with the conclusion derived for
isolated orbits\cite{Fo08} confirms that the field-dependent
chemical potential oscillations are strongly damped in 2D
compensated metals.

\section{Amplitude of the first harmonics in the LK approximation using random walk analogy}

In this section, we compute the expression for the first harmonics
amplitude $A_1^{LK}$, corresponding to the frequency $F_0$, within
the LK approximation. We will check numerically that this
approximation fits very well the amplitude $A_1$ extracted from the
spectrum computed in the previous section. We first notice that
there are an infinite number of orbits along the chain
contributing to this amplitude (see Ref. \onlinecite{Fo08}) since
a semi-classical trajectory around a successive electron and hole
pockets gives a zero frequency contribution. Therefore these zero
frequency orbits can in principle be added to any orbit which has a $F_0$
contribution to form new $F_0$-orbits by juxtaposition of continuous
trajectories. These orbits yielding
the frequency $F_0$ can be classified by their successive masses
$m_e(l)=lm_e^*+(l-1)m_h^*$ or $m_h(l)=(l-1)m_e^*+lm_h^*$, where
$l$ is a positive integer. The contribution of orbits with large
effective mass is negligible at low field or high temperature
($b/t$ small) but have a significant value in the large $b/t$
limit where all the orbits have to be taken into
account. Indeed the field- and temperature-dependent damping
factors which are defined for a given set of effective masses
$m_{e(h)}^*$ by the formula

\bb
R(m_{e(h)}^*)=\frac{2\pi^2m_{e(h)}^*t/b}{\sinh(2\pi^2m_{e(h)}^*t/b)}
\ee

are close to unity in this limit. Dingle factors

\begin{figure}
\centering
\resizebox{1\columnwidth}{!}{\includegraphics*{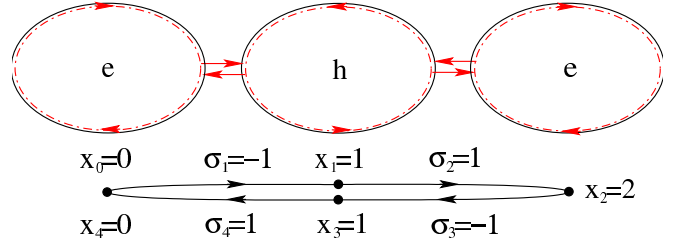}}
\caption{\label{exOrb} Example of a trajectory contributing to the
first amplitude, with fundamental frequency $F_0$ and mass
$m_e(1)=2m_e^*+m_h^*$. The successive coordinates
$\{x_i\}_{i=0,..4}$ of the quasiparticle are given below the
figure. Here $n_i=0$ (see text). $\sigma_{-1}=-1$ since the particule is going
backward when it is reaching the location $x_0=x_4$ from $x_{-1}=x_{3}$ on a
periodic orbit.}
\end{figure}

\bb \label{Dingle}
R_D(m_{e(h)}^*,t_{e(h)}^*)=\exp(-2\pi^2m_{e(h)}^*t_{e(h)}^*/b)
\ee

where $t_{e(h)}^*=m_0A_0/4\pi^2\hbar\tau_{e(h)}$ are the reduced
Dingle temperatures, can also be added to account for real
crystals in which the relaxation times $\tau_{e(h)}$ have finite
values. In the following we will assume first that these two
relaxation times are negligibly small to simplify the
calculations. The existence of an infinite set of orbits
contributing to the first harmonics leads us to define a closed
random walk on the chain in the Brillouin zone $\{x_i\}_{i=0,2n}$,
with origin and end at $x_0$ and $x_{2n}$, respectively, with the condition
$x_0=x_{2n}=0$. These coordinates take integer values (either
negative or positive) and define precisely the pocket inside which the
quasiparticle is located. We chose $x_i$ with $i$ even to be the
positions of the electron bands and $i$ odd for the hole bands. A
closed path has an even number of steps $2n$. For a given $x_i$,
the particle can also orbit a number $n_i\ge 0$ of times around
the surface before going to the next band. An example of a typical closed
trajectory is given in Fig. \ref{exOrb}.

We can rewrite coordinates $x_i$ by the mean of forward/backward
variables $\sigma_i=\pm 1$ such as
$x_{i}=x_{i-1}+\sigma_{i}(x_{i-1}-x_{i-2})$. Here $\sigma_i=1$
when the particle is going forward on its path (in the same
direction) and $\sigma_i=-1$ when it is going backward in the
reverse direction (see Fig. \ref{exOrb} for an explicit set of $\sigma_i$'s).
The path is moreover made periodic by imposing $x_i=x_{i\pm 2n}$
(and the same for variables $\sigma_i$). On Fig. \ref{exOrb},
$x_3=x_{-1}$ and the particle is moving on the reverse direction
when it is reaching the point $x_0=x_4$ along the portion of path
$(x_{-1},x_0,x_1)$, which implies that $\sigma_1=-1$.

It will be also useful in the following to introduce new periodic
variables $y_i=x_i-x_{i-1}=\pm 1$ which satisfy the simple relations
$y_i=\sigma_iy_{i-1}$ and
$\sigma_i=y_iy_{i-1}$. Let us now calculate the total number of
possible orbits with a given effective mass $m_e(l)$ (or
equivalently $m_h(l)$ for a hole orbit since the problem is
symmetric by exchange of electron and hole masses) and their
contribution to the amplitude of the first harmonic $A_1^{LK}$ with the
fundamental frequency $F_0$. If the particle is going forward,
there is an amplitude equal to $ip$ to be transmitted from one
band to the next one or $ipq$ if the particle is going backward
(it has first to perform exactly one reflection with amplitude $q$
on the band edge before being transmitted back through the
previous junction). In term of variables $\sigma_i$, this is
equivalent to write this partial amplitude as
$ip[(1+\sigma_i)+(1-\sigma_i)q ]/2=ip\sqrt{q}e^{K\sigma_i}$, with
$e^K=1/\sqrt{q}$. Then the LK amplitude corresponding to all
electron-like contributions can be written as

\bb\label{amp1}
A_1^{LK}(e)=\frac{q^2R(m_e^*)}{m_e^*}+\sum_{l\ge
2}\frac{R(m_e(l))}{m_e(l)}\times
\\ \nn
\sum_{n=1}^{2(l-1)}\frac{1}{n}\sum_{\{n_i,\sigma_i\}} (-p^2)^nq^{
n+2\sum_{j=0}^{2n-1}n_j} e^{K\sum_{i=1}^{2n}\sigma_i}
\ee

The sum over $l$ corresponds to all possible masses, and the sum
over $n\ge 1$  the number $2n$ of possible tunnellings. The first
term $l=1$ in Eq. (\ref{amp1}) is the simplest orbit of the
expansion: a closed trajectory around one electron pocket with MB
damping factor $q^2$ and effective mass $m_e^*$. The sum over $n$
is limited to $2(l-1)$ which corresponds to the extremal
trajectory, $i.e$ the particle of mass $m_e(l)$ visits a maximum
of $(l-1)$ successive electron and hole pockets outside the first
initial electron pocket before going back, $l-1$ being also the maximal
linear extension of the path. This implies that all the $n_i$ are
zero for this case (see Fig. \ref{exOrb}). The factor $1/n$ takes
into account of the orbit symmetry by circular permutation of the
coordinates $\{x_i\}$. Finally, the sum over the set
$\{n_i,\sigma_i\}$ is constrained by the boundary conditions
$x_0=x_{2n}$ and by the facts that the frequency is set to $F_0$
and the effective mass is $lm_e^*+(l-1)m_h^*$.  This implies the
following conditions on $\{n_i,\sigma_i\}$:

\bb\label{cond1}
l&=&\sum_{j=0}^{n-1}n_{2j}+\frac{n}{2}+\sum_{j=0}^{n-1}\frac{1-\sigma_{2j+1}}{4}
\\ \label{cond2}
l-1&=&\sum_{j=0}^{n-1}n_{2j+1}+\frac{n}{2}+\sum_{j=1}^{n}\frac{1-\sigma_{2j}}
{ 4 }
\ee

The constrained sum can be transformed using two Kronecker integrals around the 
complex unit circle

\bb\label{amp2}
& &A_1^{LK}(e)=\frac{q^2R(m_e^*)}{m_e^*}+\sum_{l\ge
2}\frac{R(m_e(l))}{m_e(l)}\times
\\ \nn
& &\sum_{n=1}^{2(l-1)}\frac{(-1)^n}{n}\sum_{\{n_i,\sigma_i\}}p^{2n}q^{n}
e^{K\sum_{j=1}^{2n}\sigma_j}
\oint\frac{dz}{2i\pi z}\oint\frac{dz'}{2i\pi z'}\times
\\ \nn
& &q^{2\sum_{j=0}^{n-1}n_{2j}}z^{-l+\sum_{j=0}^{n-1}n_{2j}
+\frac{n}{2}+\sum_{j=0}^{n-1}\frac{1-\sigma_{2j+1}}{4}} \times
\\ \nn
& &q^{2\sum_{j=0}^{n-1}n_{2j+1}}z'^{-l+1+\sum_{j=0}^{n-1}n_{2j+1}
+\frac{n}{2}+\sum_{j=1}^{n}\frac{1-\sigma_{2j}}{4}}
\ee

The sums over the $\{n_i=0,\dots,\infty\}$ can be performed exactly, since
$q$ is less than unity. We obtain

\bb\label{amp3}
& &A_1^{LK}(e)=\frac{q^2R(m_e^*)}{m_e^*}+\sum_{l\ge
2}\frac{R(m_e(l))}{m_e(l)}\times
\\ \nn
& &\sum_{n=1}^{2(l-1)}\frac{(-1)^n}{n}\sum_{\{\sigma_i\}}p^{2n}q^{n}
e^{K\sum_{i=1}^{2n}\sigma_i}
\oint\frac{dz}{2i\pi z}\oint\frac{dz'}{2i\pi z'}\times
\\ \nn
& &
\frac{z^{-l+3n/4}}{(1-q^2z)^n}\frac{1}{z^{\sum_{j=0}^{n-1}\sigma_{2j+1}/4} }
\frac{z'^{-l+1+3n/4}}{(1-q^2z')^n}\frac{1}{z'^{\sum_{j=1}^n\sigma_{2j}/4} }
\ee

The condition for a closed path is given by $\sum_{j=1}^{2n}
y_j=0$, with the periodic conditions $y_0=y_{2n}$, which imposes
in the previous sum the introduction of another Kronecker
function. The previous quantities depending on
$\sigma_i=y_iy_{i-1}$ can be rewritten in term of the $y_i$'s
only. Using the fact that $e^K=1/\sqrt{q}$, the sum over the terms
involving these variables is given by the quantity

\bb\nn
Z_0&=&\sum_{\{y_i=\pm
1\}}\int_0^{2\pi}\frac{d\alpha}{2\pi}e^{i\alpha\sum_{j=1}^{2n} y_j}\times
\\ \label{part}
& &\prod_{j=0}^{n-1} (zq^2)^{-y_{2j}y_{2j+1}/4}(z'q^2)^{-y_{2j+1}y_{2j+2}/4}
\ee

This is the partition function for a one dimensional periodic Ising model with
an imaginary field and alternate complex coupling. Setting the $2\times 2$
transfer matrices $P_{yy'}=(zq^2)^{-yy'/4}e^{i\alpha(y+y')/2}$ and
$P'_{yy'}=(z'q^2)^{-yy'/4}e^{i\alpha(y+y')/2}$, $Z_0$ can be written as
a trace over the product of operators $(PP')^n$:

\bb
Z_0={\rm Tr} ( PP' )^n=\lambda_+^n+\lambda_-^n
\ee

 where $\lambda_{\pm}$ are the eigenvalues of the
matrix $PP'$. The amplitude can then be written as

\bb\label{amp4}
& &A_1^{LK}(e)=\frac{q^2R(m_e^*)}{m_e^*}+\sum_{l\ge
2}\frac{R(m_e(l))}{m_e(l)}\times
\\ \nn
& &\sum_{n=1}^{2(l-1)}\frac{(-1)^n}{n}p^{2n}q^{n}
\oint\frac{dz}{2i\pi z}\oint\frac{dz'}{2i\pi z'}\times
\\ \nn
& &
\frac{z^{-l+3n/4}}{(1-q^2z)^n}
\frac{z'^{-l+1+3n/4}}{(1-q^2z')^n}
Z_0.
\ee

 It is useful to express complex vectors $z=e^{i\theta}$ and
$z'=e^{i\theta'}$ by their angles $\theta$ and $\theta'$. Also, setting
$\zeta=\sqrt{q}e^{i\theta/4}$,
$\zeta'=\sqrt{q}e^{i\theta'/4}$, $\bar\zeta=1/\zeta$ and
$\bar\zeta'=1/\zeta'$, we can express the eigenvalues as

\bb
\lambda_{\pm}&=&\nu(\theta,\theta',\alpha)\pm\sqrt{\nu^2+\eta(\theta,\theta')},
\\ \nn
\nu(\theta,\theta',\alpha)&=&\bar\zeta\bar\zeta'\cos(2\alpha)+\zeta\zeta',
\\ \nn
\eta(\theta,\theta')&=&-(\zeta^2-\bar\zeta^2)(\zeta'^2-\bar\zeta'^2)
\\ \nn
&=&-(1-q^2z)(1-q^2z')/(q^2\sqrt{zz'}).
\ee

Then

\bb
& &Z_0=2\sum_{i=0}^{n/2}\sum_{j=0}^{i}\bin{n}{2i}\bin{i}{j}
\nu^ { n-2j }\eta^{j}
\\ \nn
&=&2\sum_{i=0}^{n/2}\sum_{j=0}^{i}\sum_{k=0}^{n-2j}\bin{n}{
2i } \bin { i } { j }\bin{n-2j}{k}(\zeta\zeta')^{n-2j-2k}\eta^j\cos^k(2\alpha)
\ee

where we introduced binomial coefficients $\bin{n}{k}=n!/k!(n-k)!$.
In the last line, we can perform the integral over $\alpha$ of the integrand
$\cos^{k}(2\alpha)$. The resulting integral is not zero only for $k$ even:

\bb
\int_0^{2\pi}\frac{d\alpha}{2\pi}\cos^{2k}(2\alpha)=\frac{1}{2^{2k}}\bin{2k
} { k }.
\ee

We finally obtain

\bb\nn
Z_0=2\sum_{i=0}^{n/2}\sum_{j=0}^{i}\sum_{k=0}^{n/2-j}\bin{n
} {2i } \bin { i } {
j}\bin{n-2j}{2k}\bin{2k}{k}\frac{(\zeta\zeta')^{n-2j-4k}\eta^j}{2^{2k}}.
\ee

The amplitude $A_1^{LK}$ can be rearranged using the previous results like

\bb
& &A_1^{LK}(e)=\frac{q^2R(m_e^*)}{m_e^*}+\sum_{l\ge
2}\frac{R(m_e(l))}{m_e(l)}\times
\\ \nn
& &\sum_{n=1}^{2(l-1)}2\frac{(-1)^n}{n}p^{2n}
\sum_{i=0}^{n/2}\sum_{j=0}^{i}\sum_{k=0}^{n/2-j}
\bin{n} {2i }
\bin {i}{j}
\bin{n-2j}{2k}
\bin{2k}{k}
\frac{(-1)^j}{2^{2k}}\times
\\ \nn
& &
\oint\frac{dz}{2i\pi z}\oint\frac{dz'}{2i\pi z'}\times
\frac{z^{-l+n-j-k}}{(1-q^2z)^{n-j}}
\frac{z'^{-l+1+n-j-k}}{(1-q^2z')^{n-j}}q^{2n-4j-4k}
\ee

To compute the last two complex integrals, we use the relation for any positive
integers
$(\alpha,\beta)$

\bb
\oint\frac{dz}{2i\pi z}
\frac{z^{-\alpha}}{(1-q^2z)^{\beta}}=\bin{\beta+\alpha-1}{\alpha}q^{2\alpha}
\ee

then

\bb\nn
& &A_1^{LK}(e)=\frac{q^2R(m_e^*)}{m_e^*}+\sum_{l\ge
2}\frac{R(m_e(l))}{m_e(l)}\times
\\ \label{amp5}
& &\sum_{n=1}^{2(l-1)}(-1)^np^{2n}q^{4l-2n-2}\coef(l,n)
\ee

where we defined the following combinatorial quantities

\bb\nn
& &\coef(l,n)=\frac{2}{n}\sum_{i=0}^{n/2}\sum_{j=0}^{i}\sum_{k=0}^{n/2-j}
\frac{(-1)^j}{2^{2k}}\times
\\ \label{coeffs}
& &\bin{n} {2i }
\bin {i}{j}
\bin{n-2j}{2k}
\bin{2k}{k}
\bin{l+k-1}{l-n+j+k}
\bin{l+k-2}{l-n+j+k-1}.
\ee

These positive integers $\coef(l,n)$ count the number of
non-equivalent orbits for a given mass $m_{e(h)}(l)$ and for which
the quasiparticle is visiting $2n$ successive pockets. In table I we
have reported, for information, the first numbers for increasing values of
$l$ from 2 up to 7, using relation (\ref{coeffs}). For a given $l$, $n$ is taken
from 1 to its maximum value $2(l-1)$.

\begin{table}[ht]
\caption{First values of coefficients $\coef(l,n)$ representing
the number of non-equivalent orbits for a given mass
$m_{e(h)}(l)$ with $2n$ magnetic breakdowns, $1\le n\le 2(l-1)$. }
\centering
\begin{tabular}{| c |  l l l l l l |}
\hline
$_n\backslash^l$ & 2 & 3 & 4 & 5 & 6  & 7
\\
\hline
1 & 2 & 2 & 2 & 2 & 2 & 2
\\
2 & 1 & 9 & 23 & 43 & 69 & 101
\\
3 &  & 8 & 68 & 264 & 720 & 1600
\\
4 &  & 1 & 63 & 610 & 3080 & 10925
\\
5 &  &  & 18 & 584 &  6132 & 36980
\\
6 &  &  & 1 & 228 & 5950 & 66374
\\
7 &  &  &  & 32 & 2800 & 64952
\\
8 &  &  &  & 1  & 600 & 34550
\\
9 &  &  &  &  & 50 & 9650
\\
10 &  &  &  &  & 1 & 1305
\\
11 &  &  &  &  &  & 72
\\
12 &  &  &  &  &  & 1
\\
\hline
\end{tabular}
\end{table}

Finally, the total amplitude for the first harmonics is given by the sum of
hole and electron contributions:

\bb\label{amp6}
A_1^{LK}=\frac{F_0}{\pi}\Big [A_1^{LK}(e)+A_1^{LK}(h) \Big ].
\ee

\begin{figure}                                                      
\centering \resizebox{0.8\columnwidth}{!}
{\includegraphics*{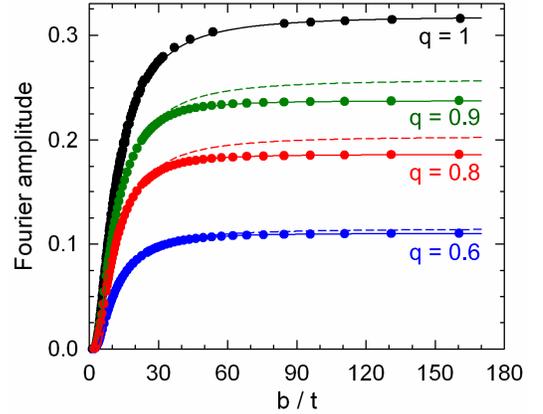}} \caption{\label{fig_h1}(color
online) $b/t$ dependence of the first harmonics amplitude. Solid
symbols are deduced from the numerical resolution of Eq.
(\ref{freeE}). Solid lines correspond to the LK approximation
given by Eqs. (\ref{amp5}) to (\ref{amp6}). Dashed lines
correspond to the first order term ($l$=1 in Eqs. (\ref{amp5}) ).
The effective masses are $m_e$=1 and $m_h$=2.5.}
\end{figure}

The first harmonics amplitude is determined by the set of Eqs.
(\ref{amp5}) to (\ref{amp6}). Examples are given in Fig.
\ref{fig_h1} (solid lines) and compared to the numerical
resolution of the Fourier spectrum of the magnetization taken from Eq.
(\ref{freeE}) (solid symbols) and for different $q$ values.
Since the $q$ is fixed, these data can be regarded
as temperature-dependent amplitudes at a given field. An excellent
agreement between numerical results and Eqs. (\ref{amp5}) to
(\ref{amp6}) is observed, even at high magnetic field. Dashed
lines in this figure are the contributions of the first order
terms (Eq. \ref{amp6} reduces to
$A_1^{LK}=q^2F_0[R(m_e^*)/m_{e}^*+R(m_h^*)/m_h^*]/\pi$ within this
approximation). These terms, which only takes into account the
basic orbits with the lowest effective masses ($m_e^*$ and
$m_h^*$) are strongly dominant since only a small difference (less
than 10 $\%$) is observed in the high $b/t$ range. However, as
discussed below, the high-order terms with higher effective masses
can have a significant influence on the evaluation of the
effective masses from experimental data.

In the case of real experimental data collected on q-2D
compensated metals, an effective mass $m^*$ is deduced from the
temperature dependence of the amplitude of the first harmonics
$A_1$ at a fixed magnetic field. In such a case, it is implicitly
assumed that either the effective mass of electron and hole orbits
is the same or, oppositely, that only one orbit contributes to the
considered Fourier component because the other has a much larger
effective mass. In addition, the contribution of high order orbits
($l>$1) is neglected. Taking into account the Dingle damping
factor (see Eq. \ref{Dingle}), we can rewrite the LK formula as:

\begin{equation}
\label{eq_LK_mean} y \propto R_{MB}
\frac{e^{-2\pi^2m^*t^*/b}}{\sinh(2\pi^2m^*t/b)}
\end{equation}

where $y=A_1b/t$, $R_{MB}$ is the relevant MB damping factor and
$t^*$ is the reduced Dingle temperature. If the magnetic field is
fixed, Eq. (\ref{eq_LK_mean}) reduces to $y\propto
1/\sinh(2\pi^2m^*t/b)$. Therefore, the effective mass can be
extracted by considering the following combination of derivatives:

\begin{equation}
\label{eq_m*} (2\pi^2m^*)^2=2\Big [\frac{1}{y}
\frac{\partial y}{\partial (t/b)} \Big ]^2 -
\frac{1}{y}
\frac{\partial^2 y}{\partial^2 (t/b)}
\end{equation}

\begin{figure}                                                      
\centering \resizebox{0.8\columnwidth}{!}
{\includegraphics*{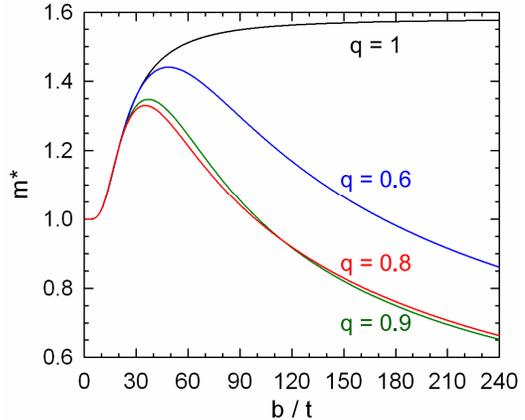}}
\caption{\label{fig_m_effective}(color online) $b/t$ dependence of
the effective mass deduced from Eq. (\ref{eq_m*}) with the
parameters relevant to data in Fig. \ref{fig_h1}. }
\end{figure}

\begin{figure}                                                      
\centering \resizebox{0.8\columnwidth}{!}
{\includegraphics*{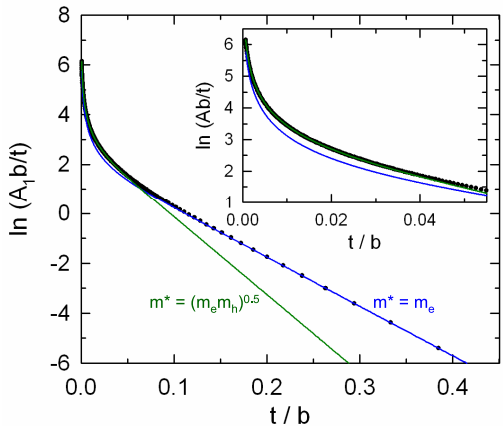}} \caption{\label{fig_lnY}(color
online) $t/b$ dependence of $y=A_1b/t$ deduced from Eqs.
(\ref{amp5}) to (\ref{amp6}) for $m^*_e$=1, $m^*_h$=2.5 and $q$=1
(solid symbols) and best fits of Eq. (\ref{eq_m*}) to this data in
the low (blue solid line) and high (green solid line) $t/b$ range.
The inset displays the low $t/b$ range.}
\end{figure}

Figure \ref{fig_m_effective} displays the $b/t$ dependence of
$m^*$ corresponding to the data in Fig. \ref{fig_h1} which stands
for a perfect crystal ($t^*$=0). For a given $q$ value, the
results are scaled as $m^*$/min($m^*_e$, $m^*_h$). For $q$=1, data
analysis based on Eq. (\ref{eq_m*}) yields $m^*$/min($m^*_e$,
$m^*_h$)=1 and $\sqrt{m^*_em^*_h}$/min($m^*_e$, $m^*_h$) in the
low and high $b/t$ limit, respectively. This point is further
supported by considering the mass plot in Fig. \ref{fig_lnY} which
yields a straight line at high $t/b$.

\begin{figure}                                                      
\centering \resizebox{0.8\columnwidth}{!}
{\includegraphics*{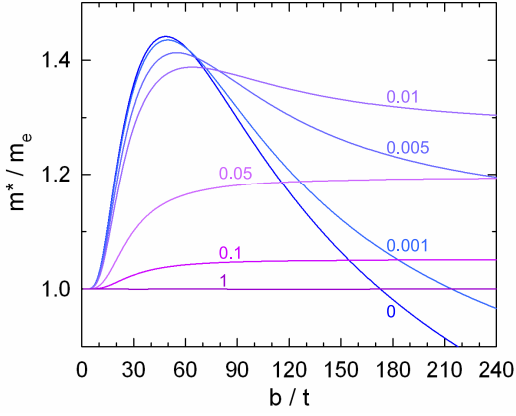}}
\caption{\label{fig_m_effective_xD}(color online) $b/t$ dependence
of the effective mass deduced from Eq. (\ref{eq_m*}) for
$m^*_e$=1, $m^*_h$=2.5, $q$=0.6 and various values of the Dingle
temperature ($t^*$) indicated on the curves. }
\end{figure}

For $q < 1$, a strongly non-monotonic $b/t$ dependence of $m^*$ is
observed in Fig. \ref{fig_m_effective}. This behavior is linked to
the high order terms ($l > 1$ in Eqs. (\ref{amp5}),
(\ref{coeffs})). The $q$-dependence of this behavior is also
strongly non-monotonic since numerous zeroes can arise in the
coefficients involved in Eq. (\ref{amp5}) as $q$ varies.
Nevertheless, the effective mass variations are damped for real
crystals with finite Dingle temperature, as demonstrated in Fig.
\ref{fig_m_effective_xD} for $q$=0.6 where it is assumed that
$t^*_e=t^*_h$ for simplicity. Assuming a fixed magnetic field of
30 T (which yields a MB field $B_0$=30.6 T for $q$=0.6),
$t^*/b$=0.01 stand for a good crystal with a Dingle temperature
$T_D$=0.4 K. Oppositely, $t^*/b$=1, for which $m^*$ is always
close to min($m^*_e$, $m^*_h$) corresponds to an extremely bad
crystal for which $\omega_c\tau$ = $b$/$2\pi m^*t^*$
(where the cyclotron frequency $\omega_c=2\pi\hbar b/m_0m^*A_0$ in our units) is
much lower than 1 at experimentally accessible fields.

\section{Properties of the second harmonics in the Canonical Ensemble}

 In this section, we study the $t/b$ dependence of the second
harmonics. If the analytical expression is difficult to obtain for
general parameter $q$, it is however possible to obtain some
information in the limiting case $q=1$, which has already been
considered in a previous study \cite{Fo08}. In particular, we
have seen that, in this limit, the free energy $\Delta F$, for a
compensated metal, is given by the difference between the Grand
potentials of the electron and hole bands, respectively:

\bb\label{Fe}
\Delta F&=&\Omega_e-\Omega_h\simeq
-m_e^*\frac{\mu^2}{2}-m_h^*\frac{(\Delta-\mu)^2}{2}
\\ \nn
&+&
\frac{b^2}{2m_e^*}
\sum_{n=1}^{\infty}\frac{(-1)^n}{\pi^2n^2}R(nm_{e}^*)\cos(2\pi
nm_e^*\frac{\mu}{b})
\\ \nn
&+&
\frac{b^2}{2m_h^*}
\sum_{n=1}^{\infty}\frac{(-1)^n}{\pi^2n^2}R(nm_{h}^*)\cos(2\pi
nm_h^*\frac{\Delta-\mu}{b}).
\ee

The chemical potential $\mu$ satisfies the self-consistent equation
$\partial \Delta F/\partial \mu=0$ given by

\bb \nn \mu&=&\EF+\frac{b}{m^*_e+m^*_h}
\sum_{n=1}^{\infty}\frac{(-1)^n}{\pi n} [ R(nm_{h}^*)\sin(2\pi
nm_h^*\frac{\Delta-\mu}{b})
\\ \label{mu}
&-&R(nm_{e}^*)\sin(2\pi nm_e^*\frac{\mu}{b}) ]. \ee

We follow the appendix $B$ of reference (\onlinecite{Fo08}) to
extract (for small field values) the second harmonics from
equations (\ref{Fe}, \ref{mu}). The oscillatory part of the
magnetization is given by $M_{osc}=-\partial \Delta F/\partial
b=x^2\partial \Delta F/\partial x$ (we remind that $x=1/b$). We
then introduce the periodic function

\bb\label{Gfunction}
G(x)=\sum_{n=1}^{\infty}\frac{(-1)^n}{\pi n}
[R(nm_{h}^*)-R(nm_{e}^*)]\sin(2\pi nF_0x)
\ee

so that the chemical potential Eq. (\ref{mu}) can be expressed as
$\mu=\EF+bG(x)/(m^*_e+m^*_h)$, and the magnetization


\bb\label{Mosc}
M_{osc}&\approx& -\frac{1}{(m_e^*+m_h^*)}G(x)G'(x)
\\ \nn
&+&\frac{1}{2}
\sum_{n=1}^{\infty}\frac{(-1)^n}{\pi^2n^2}\Re
\Big [
\\ \nn
& &\frac{R(nm_{e}^*)}{m_e^*}\frac{\partial}{\partial x} \exp(2i\pi
nF_0x+2i\pi nw_eG(x))
\\ \nn
&+&
\frac{R(nm_{h}^*)}{m_h^*}\frac{\partial}{\partial x} \exp(2i\pi
nF_0x-2i\pi nw_hG(x)
\Big ]
\ee

with $w_{e(h)}=m_{e(h)}^*/(m_e^*+m_h^*)$. We make the further approximation
in the exponential parts of (\ref{Mosc}), and which is valid at low temperature,
that $G(x)$ can be truncated to the first term
$G(x)\approx[R(m_{e}^*)-R(m_{h}^*)]\sin(2\pi F_0x)/\pi$ so that

\bb\label{bess}
e^{2i\pi nw_{e(h)}G(x)}
=
\sum_{m=-\infty}^{\infty}J_{m}(n\alpha_{e(h)})\exp(2i\pi mF_0x)
\ee

where
$\alpha_{e(h)}=2m_{e(h)}[R(m_{e}^*)-R(m_{h}^*)]/(m_e^*+m_h^*)$,
and $J_m$ are the Bessel functions of integer order. Putting
expression (\ref{bess}) inside (\ref{Mosc}), we then select, in order
to isolate the second harmonics, integers such that $n+m=\pm 2$ and $n-m=\pm
2$ for the electron and hole contributions, respectively. Expanding
the magnetization in terms of Fourier components
$M_{osc}=A_1\sin(2\pi F_0)+A_2\sin(4\pi F_0)+\dots$, we find that
the coefficient $A_2$ satisfies the relation

\bb\label{A2}
& &\frac{\pi A_2}{2F_0}=
-\frac{(R(m_h^*)-R(m_e^*))^2}{2(m_e^*+m_h^*)}+
\\ \nn
& &\sum_{n\ge 1}\frac{
[
R((n+2)m_h^*)-R((n+2)m_e^*)
][
R(n m_h^*)-R(n m_e^*)]}
{n(n+2)(m_e^*+m_h^*)}
\\ \nn
&-& \sum_{n\ge 1}\frac{1}{n^2}
\Big [
\frac{R(nm_{e}^*)}{m_e^*}(J_{n-2}(n\alpha_e)-J_{n+2}(n\alpha_e))
\\ \nn
&+& \frac{R(nm_{h}^*)}{m_h^*}(J_{n-2}(n\alpha_h)-J_{n+2}(n\alpha_h))
\Big ].
\ee

\begin{figure}
\centering
\resizebox{1.2\columnwidth}{!}{\includegraphics*[angle=270]{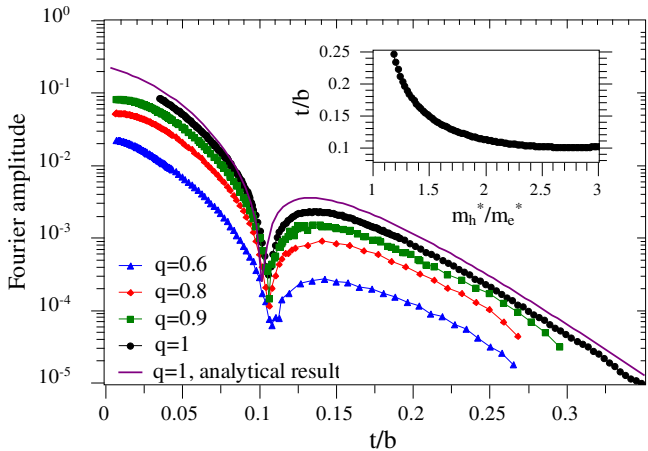}}
\caption{\label{ampA2} (color online) t/b dependence of the second
harmonics amplitude $A_2$ for various $q$ values. Solid symbols
and solid line correspond to the numerical resolution of Eq.
\ref{freeE} and to the solution of approximation (\ref{A2}) for
$q=1$, respectively. Here $m_e^*=1$ and $m_h^*=5/2$. In the insert
is plotted the critical value of $t/b$ at which $A_2$ vanishes as
function of the ratio $m_h^*/m_e^*$.}
\end{figure}

In Fig. \ref{ampA2} is plotted the expression (\ref{A2}) together
with the numerical results obtained by solving directly the
spectrum (\ref{spectrum}) and extracting the second harmonics
amplitude for different values of $q$ (see section \ref{model}).
The two results at $q=1$ agree quite well for a large domain of
$t/b$. Remarkably, the numerical data show that, whatever the $q$
value is, the amplitude vanishes at a unique value of $t/b$
depending on the ratio of the two effective masses only. In the
example given by $m_e^*=1$ and $m_h^*=5/2$, the amplitude vanishes
for $t/b\simeq 0.1$. This behavior is strikingly different from
that of $A_1=A_1^{LK}$, considered in previous section, for which
no singularity is observed. This reflects the fact that the
deviation from the LK approximation appears only in the second
harmonic.

\section{Summary and conclusion}

The spectrum for one-dimensional chain of compensated orbits has been
calculated. As it is the case for two isolated orbits, the
field-dependent oscillations of the chemical potential are weak. In turns, the
LK formalism can be applied,
provided MB orbits, which in such a network contributes to the
fundamental Fourier component amplitude, although with higher
effective masses, are taken into account. The resulting high order
terms ($l >$ 1 in Eqs. (\ref{amp5}) and (\ref{coeffs})) can lead
to apparent temperature-dependent effective mass for clean
crystals in the high $b/t$ limit in the case where only one
effective mass is considered for the data analysis, as it is
usually done. On the contrary, strong deviation from the LK
behavior is observed for the second harmonics. The main feature of
this latter component being the zero amplitude occurring at a $t/b$
value depending only on the effective mass ratio $m^*_h/m^*_e$.
Finally, it can be remarked that the studied 1D-chain does not yield
frequency combinations, only harmonics of the fundamental
frequency. In a next step, it is planned to consider 2D networks
of compensated orbits which account for the FS of many organic
metals and are known to give rise to such phenomenon.

\bibliographystyle{unsrt}
\bibliography{1D_chain}

\end{document}